\documentclass[runningheads]{llncs}
\usepackage[T1]{fontenc}
\usepackage{graphicx}
\usepackage[utf8]{inputenc} %
\usepackage[T1]{fontenc}    %
\usepackage[colorlinks,allcolors=black]{hyperref}       %
\usepackage{url}            %
\usepackage{booktabs}       %
\usepackage{amsfonts}       %
\usepackage{nicefrac}       %
\usepackage{microtype}      %
\usepackage{xcolor}         %
\usepackage{latexsym}
\usepackage{booktabs}
\usepackage{enumitem}
\usepackage{subfigure}
\usepackage{color}
\usepackage{comment}
 \usepackage{amsmath} 
\usepackage{tikz}
\usepackage{graphicx}
\usepackage{todonotes}
\usetikzlibrary{matrix,fadings,arrows,trees,calc,positioning,decorations,automata}
\usepackage{circledsteps}
\usepackage{bbding} %

\begin{document}
\title{Prompt Compression in the Wild: Measuring Latency, Rate Adherence, and Quality for Faster LLM Inference}
\titlerunning{Prompt Compression in the Wild}
\author{Cornelius Kummer \orcidID{00000-0002-5248-4532} \and
Lena Jurkschat \Envelope \orcidID{0009-0002-7332-5861} \and
Michael Färber \orcidID{0000-0001-5458-8645} \and
Sahar Vahdati \orcidID{0000-0002-7171-169X}}

\authorrunning{C. Kummer et al.}
\institute{ScaDS.AI Dresden/Leipzig, CIDS, TU Dresden\\
\email{firstname.lastname@tu-dresden.de}}
\maketitle              %
\begin{abstract}
With the wide adoption of language models for IR -- and specifically RAG systems -- the latency of the underlying LLM becomes a crucial bottleneck, since the long contexts of retrieved passages lead large prompts and therefore, compute increase.
Prompt compression, which reduces the size of input prompts while aiming to preserve performance on downstream tasks, has established itself as a cost-effective and low-latency method for accelerating inference in large language models. However, its usefulness depends on whether the additional preprocessing time during generation is offset by faster decoding. We present the first systematic, large-scale study of this trade-off, with thousands of runs and 30{,}000 queries across several open-source LLMs and three GPU classes. Our evaluation separates compression overhead from decoding latency while tracking output quality and memory usage. \textsc{LLMLingua} achieves up to 18\% end-to-end speed-ups, when prompt length, compression ratio, and hardware capacity are well matched, with response quality remaining statistically unchanged across summarization, code generation, and question answering tasks. Outside this operating window, however, the compression step dominates and cancels out the gains. We also show that effective compression can reduce memory usage enough to offload workloads from data center GPUs to commodity cards, with only a 0.3s increase in latency. Our open-source profiler predicts the latency break-even point for each model--hardware setup, providing practical guidance on when prompt compression delivers real-world benefits.

\end{abstract}

\section{Introduction}
\label{sec:introduction}
Large language models (LLMs) are increasingly used in information retrieval (IR) systems, for instance as rerankers, retrievers, or within retrieval-augmented generation (RAG).
RAG extends the knowledge capacity of LLMs by integrating external retrieval into the generation process.
For interactive applications such as chatbots and search assistants, low response latency is crucial.
In practice, RAG prompts often contain multiple retrieved passages, resulting in long contexts.
Efficiently processing these long prompts is therefore essential for scalable IR systems, with LLM inference latency emerging as a major bottleneck in addition to retrieval latency.  

Prompt compression addresses this challenge by reducing the prompt length while preserving task-relevant information.
The goal is to increase the information density of the input by removing redundant or irrelevant tokens, such that the generated output remains comparable in quality and semantics to the original.
A prominent family of approaches is \textsc{LLMLingua}, which prunes tokens either via perplexity-based importance estimation using a small language model, or via token-level classification with encoder models \cite{jiang-etal-2023-llmlingua,jiang-etal-2024-longllmlingua,pan-etal-2024-llmlingua}.
As text-to-text methods, these compressors are model-agnostic and compatible with black-box APIs. Reported gains are substantial -- up to 5.7×, 2.6×, and 2.9× for \emph{LLMLingua} \cite{jiang-etal-2023-llmlingua}, \emph{LongLLMLingua} \cite{jiang-etal-2024-longllmlingua}, and \emph{LLMLingua-2} \cite{pan-etal-2024-llmlingua}, respectively.
However, evaluation details are sparse, and more broadly, existing work emphasizes response quality while paying little attention to latency, memory, and cost impacts \cite{nagle_fundamental_2024,selective_context_2023,jha_characterizing_2024}.  

In practice, prompt compression must balance three interdependent factors: \emph{latency}, \emph{output quality}, and \emph{cost}.
These are jointly influenced by the characteristics of the input prompt, the choice of compressor and target LLM, and the underlying hardware/software configuration.
A fair assessment of prompt compression therefore requires a comprehensive, multi-dimensional evaluation that disentangles compression overhead from actual decoding.  

In this work, we present a large-scale study of prompt compression under realistic deployment conditions.
We conduct more than 30,000 end-to-end inference experiments across five open-source model setups (7B–70B) and two proprietary APIs, on three hardware classes (Nvidia A100, GTX 1080 Ti, Apple M1 Pro).
Prompt lengths range from 100 to 50,000 tokens and compression ratios from 1.5$\times$ to 5$\times$.
Our experiments (i) isolate compressor overhead by timing it separately from decoding, (ii) quantify end-to-end latency across models, hardware, and compression ratios, (iii) verify response quality on summarization, code generation, and QA tasks, and (iv) measure memory impact, showing how compression reduces peak GPU usage and enables long-context inference on consumer-grade devices.
Overall, we find that prompt compression can reduce latency by up to \textit{18\%} without quality loss and cut GPU memory consumption by up to \textit{75\%}, with the largest gains beyond $\sim$5k-token prompts.  

\noindent Our contributions are threefold:  
\begin{itemize}[leftmargin=*]
  \item We design a unified evaluation framework that disentangles compression overhead from decoding and aligns latency, quality, and memory metrics across models, prompts, and hardware.  
  \item We run over 30,000 end-to-end experiments, demonstrating up to 18\% faster inference and 75\% lower GPU memory once prompts exceed 5k tokens.  
  \item We release an open-source profiler that predicts, for any model–GPU pair, the prompt length at which compression yields net benefit.  
\end{itemize}

We review related work in Section \ref{sec:related_work}, present prompt compression preliminaries in Section \ref{sec:preliminaries}, describe our experiments and results in Section \ref{sec:evaluation}, and conclude in Section \ref{sec:conclusion}.

\section{Related Work}
\label{sec:related_work}
\subsubsection{Acceleration for LLM-Based IR.}
LLM inference acceleration is increasingly critical as model sizes grow. In IR, users expect low-latency responses.
Prompt/input compression complements output restructuring or organising methods for faster inference \cite{zhou_survey_2024,ning2024skeletonofthoughtpromptingllmsefficient} and has been explored in LLM-based IR, including reference compression \cite{zhu_irllmsurvey_2025}.
PE-Rank compresses multi-passage inputs for re-ranking into special passage tokens \cite{liu_compressionforreranking_2025}. Other ways to shorten prompts include context filtering \cite{wang2023learningfiltercontextretrievalaugmented}, summarization/semantic compression \cite{liu-etal-2023-tcra}, and projecting document embeddings into a single token in the model's representation space \cite{cheng_xrag_2025}.
This work quantifies the latency gains achievable with LLMLingua-based prompt compression and how they vary with hardware, model, and software settings.

\subsubsection{Prompt Compression: Soft vs.\ Hard.}
Soft compression encodes prompts into compact continuous vectors with trained encoders \cite{soft_compression_adapting,soft_compression_adaptingllmsefficientcontext,soft_compression_incontextautoencodercontextcompression}.
While an effective intra-model approach, it is unsuitable for black-box LLMs lacking embedding access.
We therefore do not consider it in this work.
Hard compression operates on text—abstractive, extractive, or token-pruning \cite{jha_characterizing_2024}.
We study LLMLingua \cite{pan-etal-2024-llmlingua,jiang-etal-2023-llmlingua,jiang-etal-2024-longllmlingua}, a representative token-pruning method that preserves key semantics (even if the compressed text is ungrammatical).
Alternatives include SelectiveContext \cite{selective_context_2023}, PCRL \cite{pcrl_2024}, RECOMP \cite{recomp_2023}, and Semantic Compression \cite{semantic_comp_2024}.

\subsubsection{Latency and System Factors.}
Despite substantial progress in prompt compression, most prior studies emphasize output quality, while latency and system-level effects remain underexplored.
Nagle et al.\ \cite{nagle_fundamental_2024} formalized prompt compression as a rate–distortion problem, highlighting the trade-off between compression and quality.
Jha et al.\ \cite{jha_characterizing_2024} systematically compared methods on long-context tasks, examining performance degradation across rates.
Li et al.\ \cite{selective_context_2023} proposed SelectiveContext to reduce memory and latency, but without evaluation across hardware, software, and model variants.
We address this gap by focusing on latency in hard prompt compression for self-hosted, open-source models, where runtime efficiency is particularly critical.

\section{Preliminaries}
\label{sec:preliminaries}
\subsubsection{Prompt Compression.} 
Prompt compression is the process of transforming an input prompt \( p \) into a compressed prompt \( p' \), such that the number of tokens in \( p' \) is significantly smaller than in \( p \) (i.e., \( |p'| \ll |p| \)), while preserving the essential task-relevant information needed for the target model \( M \) to generate an output \( y \) that is semantically and qualitatively equivalent to the output produced from the original prompt.
Formally, prompt compression seeks a mapping \( f: p \mapsto p' \) that minimizes \( |p'| \) under the constraint:
$\text{Quality}(M(p')) \approx \text{Quality}(M(p))$,
where \( \text{Quality}(\cdot) \) measures semantic fidelity, task accuracy, or another task-specific evaluation metric.
\begin{table*}[tb]
    \caption{Overview of LLMLingua Prompt Compressors and Hardware benchmarked in our work}
    \label{tab:overview-compression-models}
    \centering
    \small
    \begin{tabular}{|p{2.9cm}|l|p{1cm}|p{1.4cm}|p{1.5cm}|p{1.5cm}|}
    \hline
         \textbf{LLMLingua Variant} & \textbf{Compression Model} & \textbf{Model Size} &\textbf{Nvidia A100}& \textbf{GTX 1080 Ti} & \textbf{Apple M1 Pro} \\
         \hline
         LLMLingua      & LLaMA 2 7B    & 7B    & \checkmark  & $\leq 4K$  & x \\
         LLMLingua-small & GPT-2 Small   & 124M  & \checkmark  & \checkmark   & $\leq4K$ \\
         \hline
         LLMLingua-2    & XLM-RoBERTa Large & 355M & \checkmark & \checkmark & \checkmark \\
         LLMLingua-2-small & BERT Base & 110M & \checkmark & \checkmark & \checkmark \\
         \hline
    \end{tabular}
\end{table*}

\subsubsection{LLMLingua and LLMLingua-2.} LLMLingua \cite{pan-etal-2024-llmlingua} is a compression technique, which reduces the  prompt length via token-level pruning using a small decoder language model, assuming that tokens with low information entropy  can be removed without loss of essential information.
The approach uses the perplexity metric to decide which tokens remain in the prompt.
It evaluates how well a probabilistic model predicts a sequence, calculating the exponentiated averaged negative log-likelihood of all tokens in a sequence, reflecting the certainty of a model's prediction. 
The negative log-likelihood of a token is also called \emph{self-information}.
LLMLingua-2 \cite{jiang-etal-2023-llmlingua}, on the other hand, appends a linear classification layer to an encoder model and subsequently fine-tunes it for compression using cross-entropy loss.

Naturally, both methods are lossy compression approaches, impacting the target model's downstream task performance.
Throughout this work, we use the terms compression rate and compression ratio as described in Pan et al. \cite{pan-etal-2024-llmlingua} and Jiang et al. \cite{jiang-etal-2023-llmlingua,jiang-etal-2024-longllmlingua}. The compression rate is defined as $\tau=\widetilde{L}/L$, where $L$ is the original prompt size (i.e. $L=|p|$) and $\widetilde{L}$ the target prompt size.
The compression ratio is defined as $1/\tau$.

\section{Evaluation}
\label{sec:evaluation}
The primary goal of our benchmark is to address the following research questions:
\begin{enumerate}
    \item[Q1] What LLMLingua prompt compression settings most effectively reduce LLM inference latency?
    \item[Q2] Is the downstream-task performance based on compressed prompts sufficient under the achieved inference speed-up, especially when open-source models are used?
    \item[Q3] Does the compression algorithm reliably adhere to its target rate, resulting in predictable prompt lengths? 
    \item[Q4] Which GPU memory requirements have to be met for different compression models? 
\end{enumerate}
We examined the two LLMLingua variants, each with two models of different sizes (Table \ref{tab:overview-compression-models}), across three different acceleration hardware architectures, from high-end HPC GPUs to consumer hardware.
We note that LongLLMLingua \cite{jiang-etal-2024-longllmlingua} was also analysed but not considered for this work, as the results largely follow the presented LLMLingua results.
The following hardware was used for our analysis:
\Circled{1} \textit{Nvidia A100 GPU (40\,GB)} as part of an HPC cluster including 8 A100 GPUs, 2 AMD EPYC 7352 CPUs (24 cores), and 1\,TB of RAM per node. 
\Circled{2} \textit{Nvidia GeForce GTX 1080 Ti node} with 11\,GB of VRAM per GPU and an Intel Xeon E5-2603 v4 CPU (6 cores), 100\,GB of RAM per node.
The node contains three GPUs.
\Circled{3} \textit{Apple M1 Pro} as a consumer-grade SoC platform with  8 CPU and 14 GPU cores, and 16\,GB of unified memory. 
Table \ref{tab:overview-compression-models} gives an overview of which compressors were evaluated on which GPU and up to which prompt length, mainly restricted by the GPUs memory.
Both compression techniques are evaluated using regular and small-sized models to assess the impact of compressor size on latency and to ensure feasibility on a consumer-grade MacBook Pro with an M1 Pro processor.
For the target models, we chose the 7B parameter model from Mistral \cite{jiang2023mistral7b} as well as the 70B LLaMA 3.1 model \cite{llama3herdofmodels}.
Both models were served either locally using the inference frameworks Hugging Face Transformers (HF-TF) \cite{wolf-etal-2020-transformers} or vLLM \cite{vllm:pagedattention}, or were accessed through an API persistent model server, both hosted on our Nvidia A100 GPU HPC cluster (see above) through the Hugging Face TGI framework \cite{hugging_face_tgi_2023}. 
In addition, we compare the open source models against the proprietary API accessible models \emph{GPT-3.5 Turbo} and \emph{GPT-4o mini}.
For setting the input length of the uncompressed prompt and the determination of the compressed token count, we used the \emph{tiktoken} GPT-3.5 encoder\footnote{\url{https://github.com/openai/tiktoken}}, following Pan et al. \cite{pan-etal-2024-llmlingua} for all our experiments.
Each compression measurement was repeated 12 times, of which the first 4 repetitions served as a warm-up and were omitted in the reported results.

\subsection{Compression Overhead (Microbenchmark)}
To isolate the overhead introduced by the compression process itself, we benchmarked the runtime of the compression algorithm, separating total execution time from model inference time on three hardware setups (see Table~\ref{tab:overview-compression-models}).
Compression was performed with a batch size of one using a single prompt from the \emph{LongBench} dataset \cite{bai_longbench_2024}, truncated to the target input length by randomly sampling from its $\sim$51,000-token context.
We evaluated prompt lengths ranging from 50 to 48,000 tokens and applied compression ratios of $1.5\times$, $2\times$, $3\times$, and $5\times$.
Unless otherwise specified, results refer to $2\times$ compression (i.e., 50\% compression).
Our analysis examines key latency factors: prompt length, compression ratio, and hardware configuration.
\begin{figure}[tb]
    \centering
    \subfigure[Compression-ratio-dependent compression latency for LLMLingua]{\includegraphics[width=0.52\linewidth, trim = 0 -4.5cm 0 0]{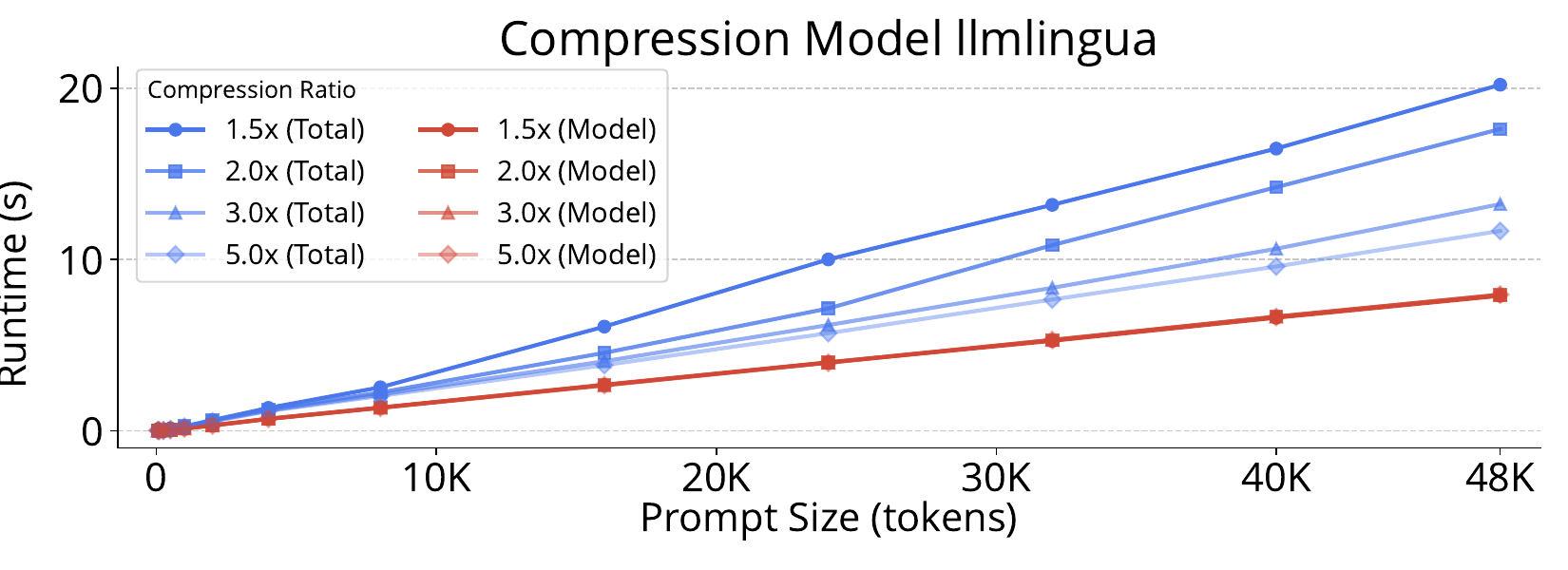}}
    \label{fig:comp-latency-hardware}\subfigure[Hardware-dependent compression latency for both LLMLingua-2 variants]{\includegraphics[width=0.47\linewidth]{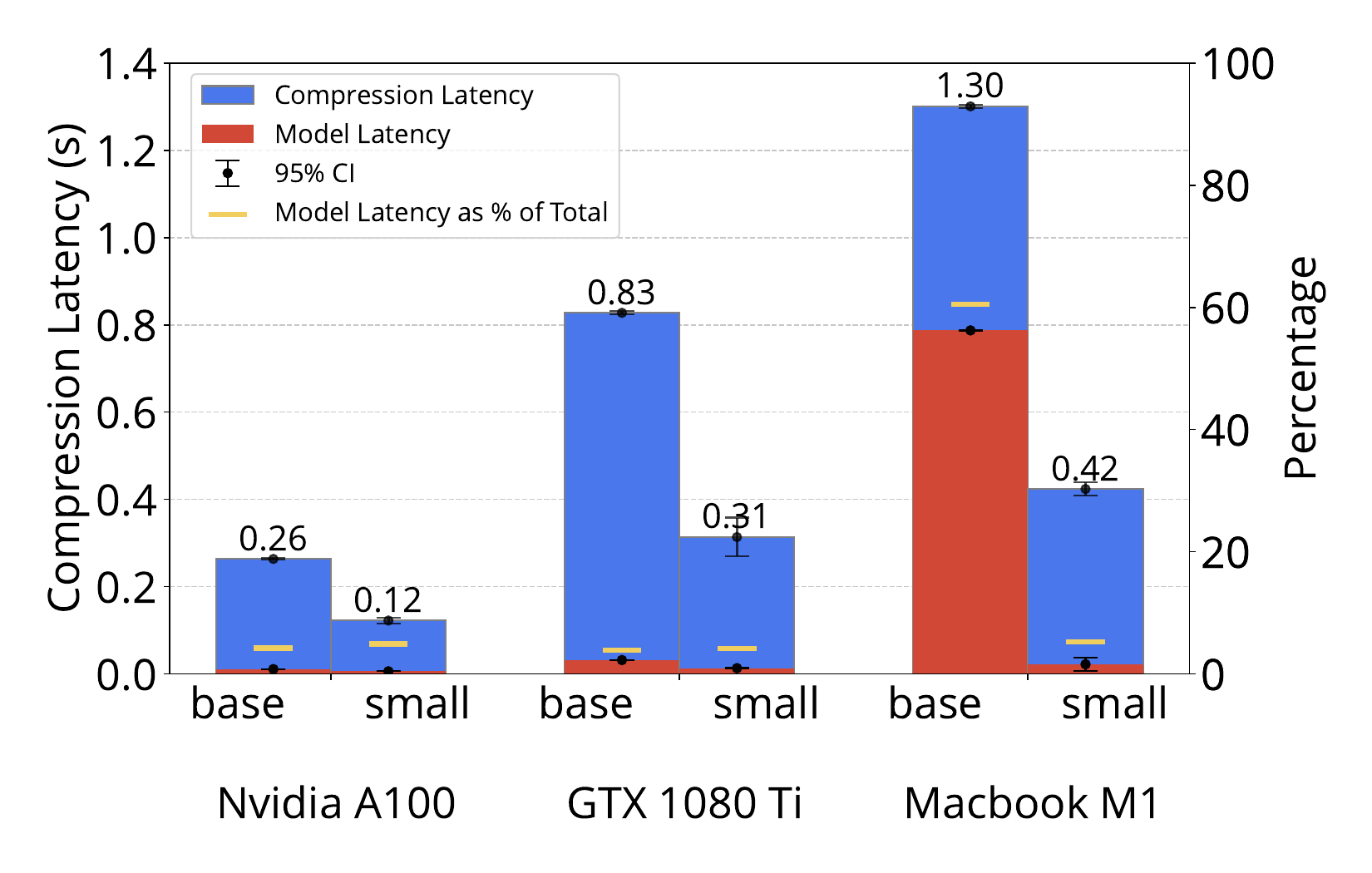}}
    \caption{Compression latency in dependence on the compression ratio (a) and on the executing hardware (b). 
    With LLMLingua-2, latency is independent of the compression ratio and reduced to max. $\sim$3s for the longest possible prompts ($48K$).
    Compression latency and model inference time percentage of LLMLingua-2 (left) and the LLMLingua-2-small variant (right) in dependence of compression hardware for a prompt size of 4,000 tokens.}
    \label{fig:comp-latency-ratio}
\end{figure}
First, we analyse the effect of the compression ratio on compression latency for LLMLingua and LLMLingua-2 on an Nvidia A100 GPU (Figure~\ref{fig:comp-latency-ratio}a).
Due to its smaller model size, LLMLingua-2 is approximately $7\times$ faster than LLMLingua.
For LLMLingua, latency decreases with higher compression ratios at a fixed prompt length, as fewer tokens are retained for iterative processing.
In contrast, LLMLingua-2's latency remains nearly constant across compression ratios, since its compression logic is not affected by the number of remaining tokens.
This independence allows compression ratio selection to focus solely on downstream performance and prompt length, without incurring additional latency costs.
For both models, total latency scales linearly with prompt size.
LLMLingua exhibits a significant overhead from the algorithmic steps surrounding model inference, growing to 13\,s at 48,000 tokens and a $1.5\times$ compression ratio, with total latency reaching 21\,s. This bottleneck stems from the fully sequential design of the algorithm.
LLMLingua-2 reduces this overhead to under 3\,s, with model inference time remaining constant between 0.01\,s and 0.03\,s, due to its direct token selection mechanism.
Small model variants show similar behaviour, but achieve faster runtimes—approximately $6\times$ faster for LLMLingua-small and $2\times$ for LLMLingua-2-small compared to their respective full models.

We further analysed the impact of hardware on compression latency, using LLMLingua-2 and its small variant with a fixed prompt length of 4,000 tokens (Figure~\ref{fig:comp-latency-ratio}b).
Switching from a high-end Nvidia A100 to a consumer-grade GTX 1080 GPU increases total latency by a factor of approximately $3\times$, and by $2\times$ for shorter prompts ($<4{,}000$ tokens).
On all setups, model inference accounts for only about 5\% of total latency.
On a MacBook M1 Pro processor, the base model’s latency increased by a factor of $5\times$, while the small variant saw a $3$–$4\times$ increase.
Notably, the model runtime constituted up to 60\% of the total latency on the M1 Pro for the base variant.
These findings highlight hardware choice as a critical factor in the real-world feasibility of compression-based acceleration.

\begin{figure}[ht!]
    \centering
    \subfigure[LLMLingua-small]{\includegraphics[width=0.7\linewidth]{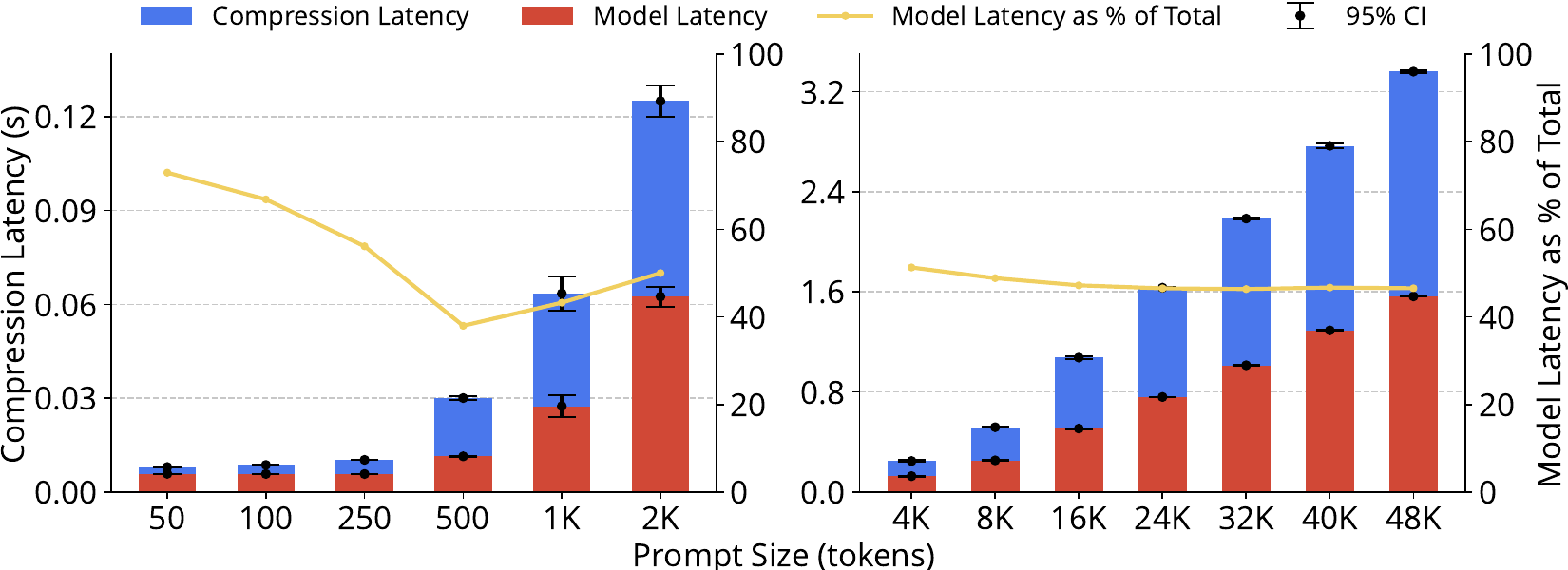}} 
    \subfigure[LLMLingua-2-small]{\includegraphics[width=0.7\linewidth]{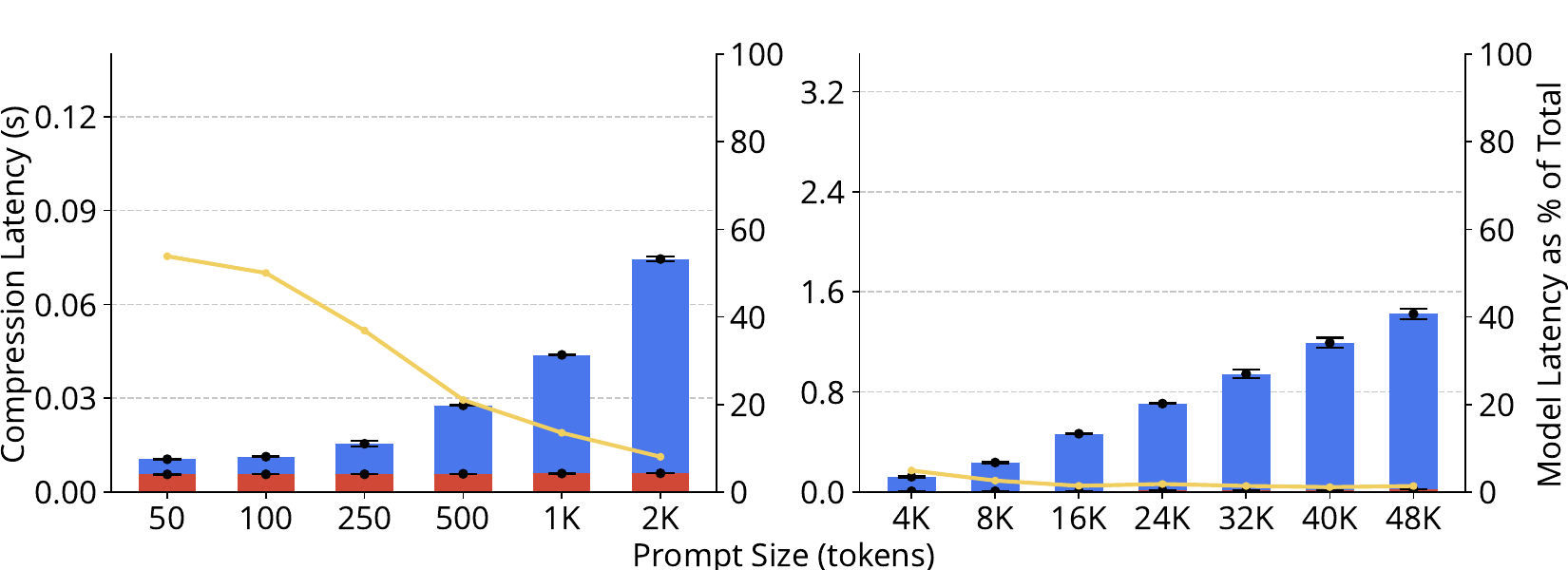}}
        \vspace{-0.4cm}
    \caption{Total prompt compression latency of the small LLMLingua variants under increasing prompt size, using a compression rate of $0.5$ and an Nvidia A100 GPU. Model inference latency as a percentage of the overall compression latency is shown in yellow.}
    \label{fig:comp-latency-promt-len}
\end{figure}

Finally, we investigated the effect of prompt size on latency for the small model variants (Figure~\ref{fig:comp-latency-promt-len}).
As with the base models (Figure~\ref{fig:comp-latency-ratio}), total latency scales linearly with input length. For prompts $\leq 250$ tokens, both compressors exhibit similar latency.
For longer inputs, LLMLingua-2-small consistently outperforms LLMLingua-small by up to 1.5\,s, primarily due to its lower and constant model runtime regardless of input length.
Among the three examined factors—compression ratio, hardware, and prompt size—LLMLingua-2 emerges as the most efficient compressor, exhibiting the lowest latencies on all hardware setups.
Nevertheless, the compression algorithm introduces substantial overhead beyond model inference, due to the sequential post-processing of the token chunks after the compression model's forward pass.
This underscores its role as a key performance bottleneck.

    \begin{figure*}[tb]
        \centering
        \includegraphics[width=1\linewidth]{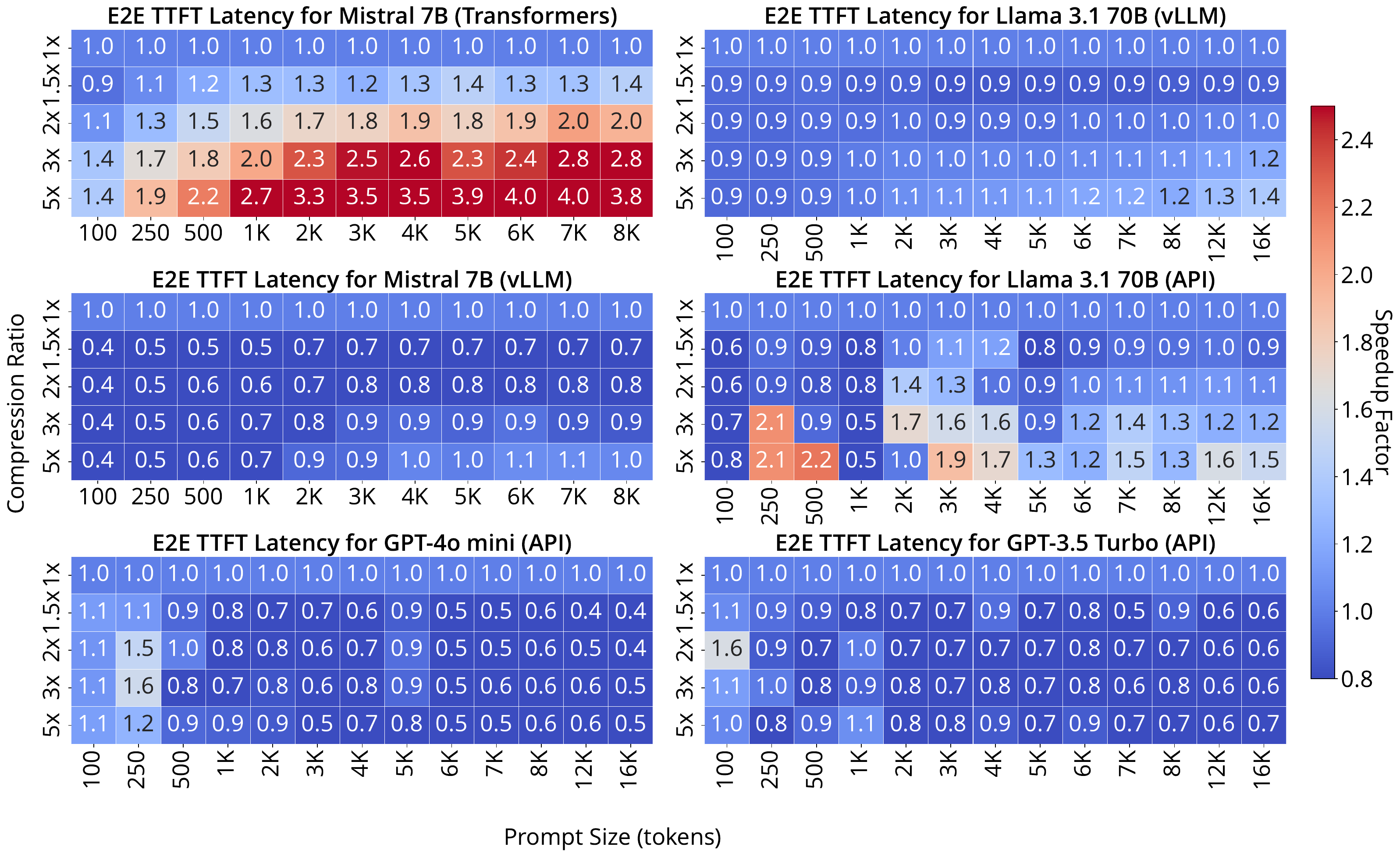}
        \caption{Speed-up for the generation of a single token (Time to First Token) for all tested target models under prompt compression with LLMLingua-2 on an Nvidia A100 GPU. The compression ratio of 1 marks the baseline, meaning no compression was applied to the prompt.}
        \label{fig:e2e-ttft}
    \end{figure*}
\subsection{End-to-End Inference Latency}
The central question in latency evaluation is whether prompt compression can accelerate the full inference process.
We assessed end-to-end (E2E) latency across multiple target LLMs differing in size and deployment strategy.
Prompt lengths ranged from 100 to 16,000 tokens, with answer lengths between 1 and 100 tokens.
Compression ratios followed those in previous latency experiments, extended by a 
baseline without compression.
To isolate the effect of prompt compression, we focused on the generation of a single token -- the Time to First Token (TTFT) -- capturing the model's prefill phase, where compression has the highest potential impact and shows the upper limit for E2E latency reductions using compression.
Therefore, Figure~\ref{fig:e2e-ttft} presents E2E speed-up for LLMLingua-2 across different compression ratios, target models, and serving frameworks. 
All experiments were run with the compression and target model on an individual Nvidia A100 GPU (40\,GB VRAM) using a batch size of one.

For smaller LLMs like Mistral 7B, observed speed-ups strongly depend on the serving framework.
Using Hugging Face Transformers, speed-ups of $3$–$4\times$ were observed for prompts exceeding 2,000 tokens and compression ratios $\geq 5\times$.
Lower ratios or shorter inputs yielded smaller gains ($\sim$2.8$\times$).
In contrast, under vLLM, speed-ups diminished significantly, with compression introducing overhead for inputs $\leq$ 4,000 tokens and compression ratios $\leq 5\times$.
These findings indicate that prompt compression offers limited benefits when models are served through optimized frameworks.
We also note, that this speed-up behaviour remains constant for longer output lengths, since the decoding phase latency is not affected by the compression.
For larger models, greater acceleration potential exists due to increased compute demand.
For LLaMA 3.1 70B served via vLLM, we observed speed-ups up to $1.4\times$ at 16k-token prompts.
However, for short inputs ($\leq 1{,}000$ tokens), compression caused overhead, reducing speed-ups to $0.9$–$1.0\times$.
Hosting the same model via TGI yielded slightly better results (up to $2\times$), likely due to reduced communication overhead, though performance variance increased due to network latency (std.\ dev.\ $0.17$ vs.\ $0.03$ for vLLM).
Commercial APIs such as OpenAI’s GPT-3.5 Turbo and GPT-4o mini showed no reliable speed-up.
Any variation was attributable to network latency rather than compression.
However, the lack of transparency and the limited control allow for no further insights on the latency variability.
For long prompts, compression often degraded performance, with speed-ups dropping below $0.5\times$.
This suggests highly optimized prefill pipelines in commercial deployments, rendering external compression ineffective.

In summary, end-to-end latency improvements through prompt compression are only observed under specific conditions: unoptimized serving frameworks (e.g., Transformers), high compression ratios, and short response lengths.
In all other settings -- especially with vLLM or commercial APIs -- the compression overhead outweighs any benefits.
Prompt compression is thus limited to narrow use cases where answer quality remains acceptable under aggressive compression and optimized inference is not available.
We also conclude, that these results are generalizable to other token pruning prompt compression methods, as the potential speed-ups are often neglected by the optimized frameworks, restricted to the prefill phase as well as they can solely diminish with increasing output lengths.

\begin{figure*}[tb]
    \subfigure[Code Generation]{\includegraphics[width=0.5\linewidth]{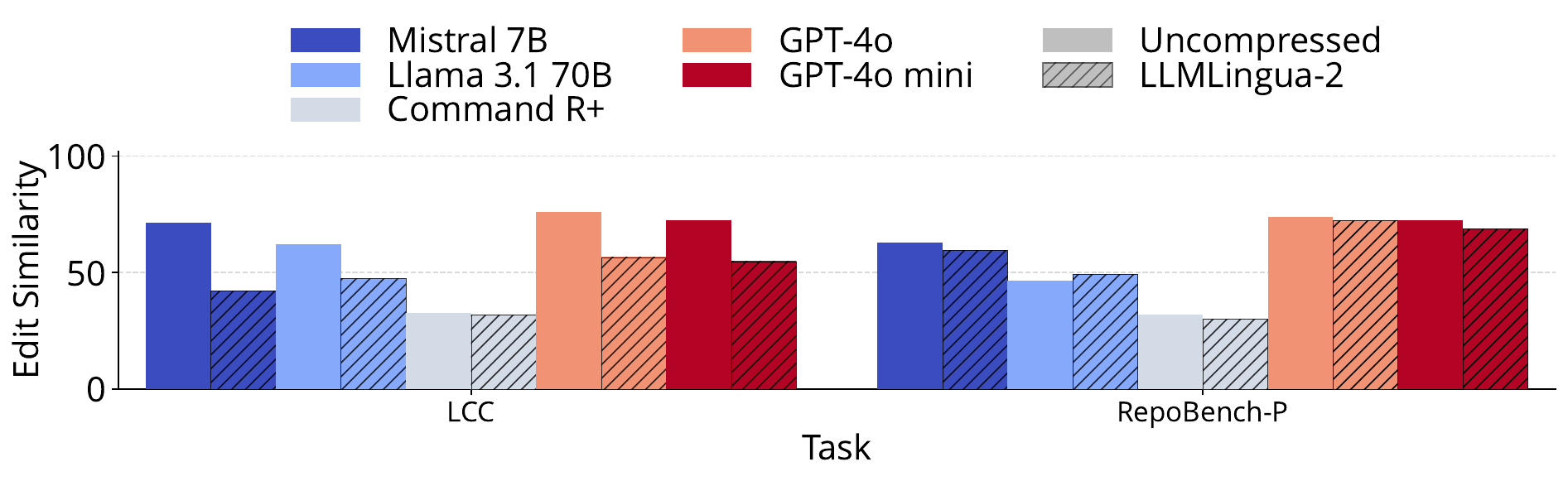}}
    \subfigure[Single-Doc QA]{\includegraphics[width=0.5\linewidth]{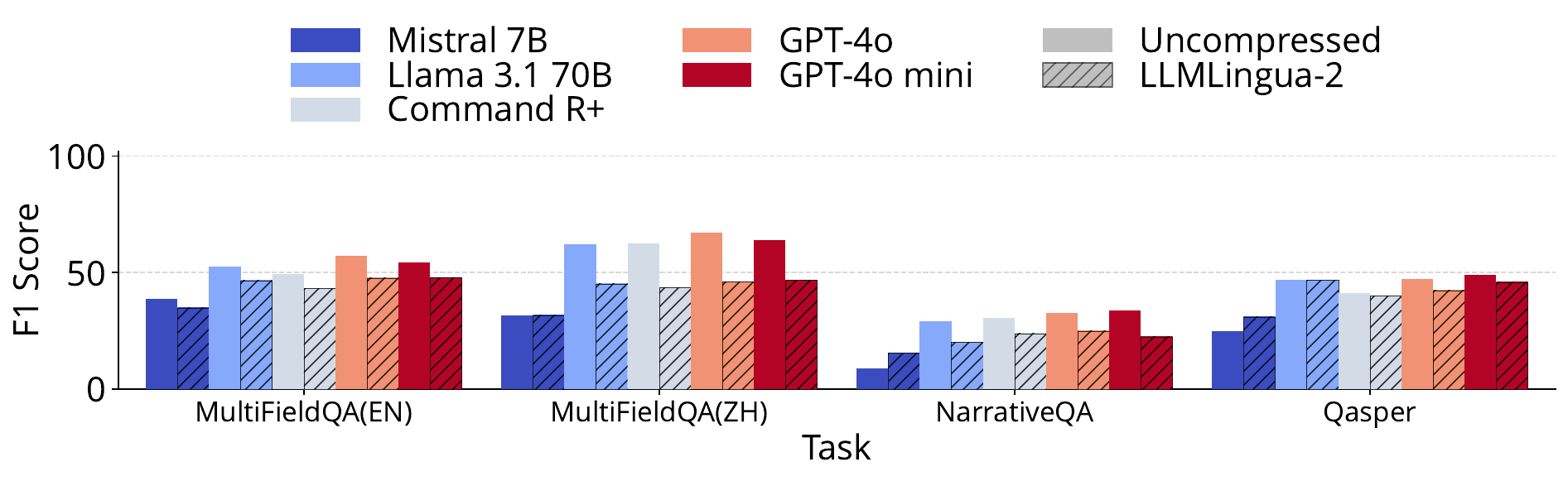}}
    \subfigure[Multi-Doc QA]{\includegraphics[width=0.5\linewidth]{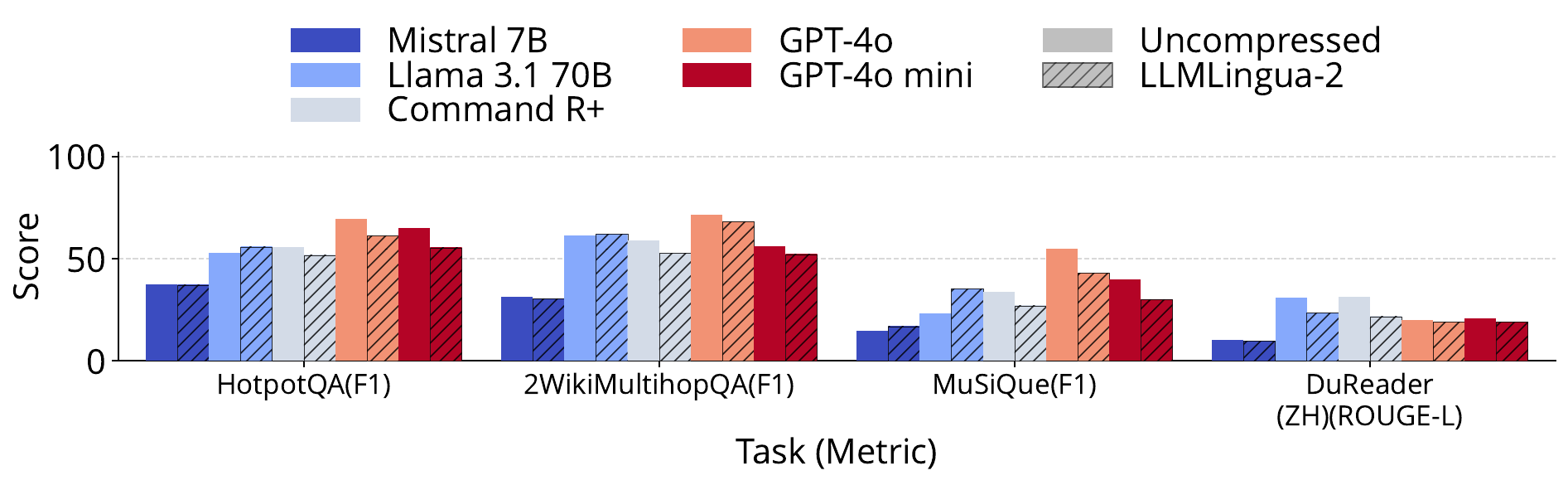}}
    \subfigure[Synthetic Tasks]{\includegraphics[width=0.5\linewidth]{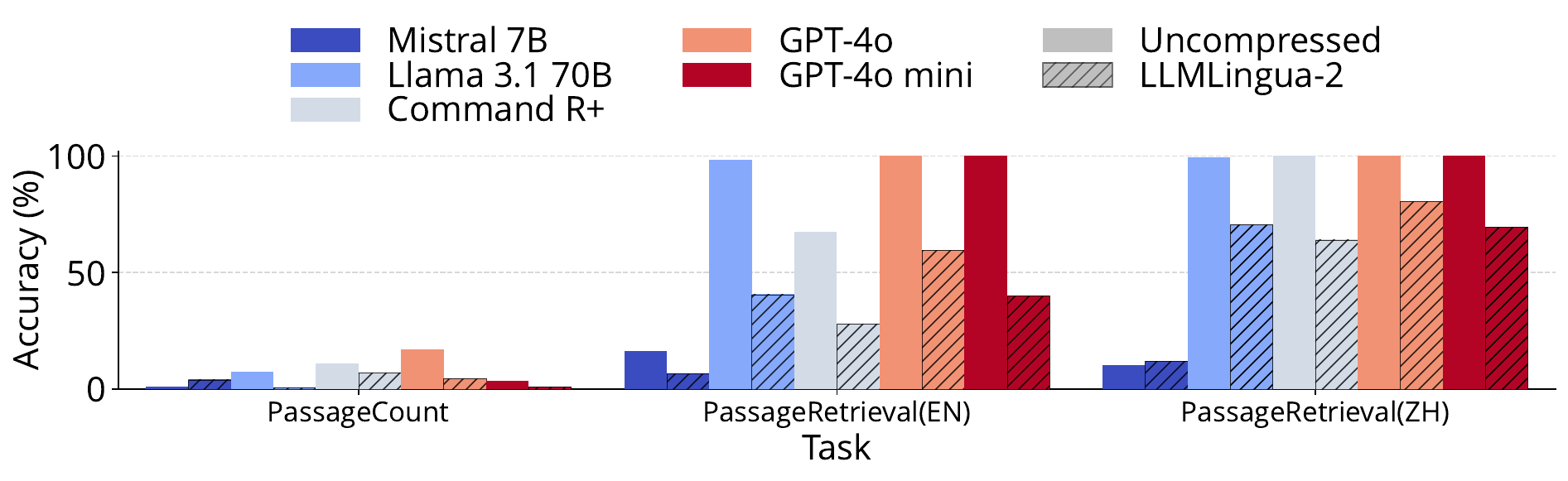}}
    \subfigure[Summarization]{\includegraphics[width=0.5\linewidth]{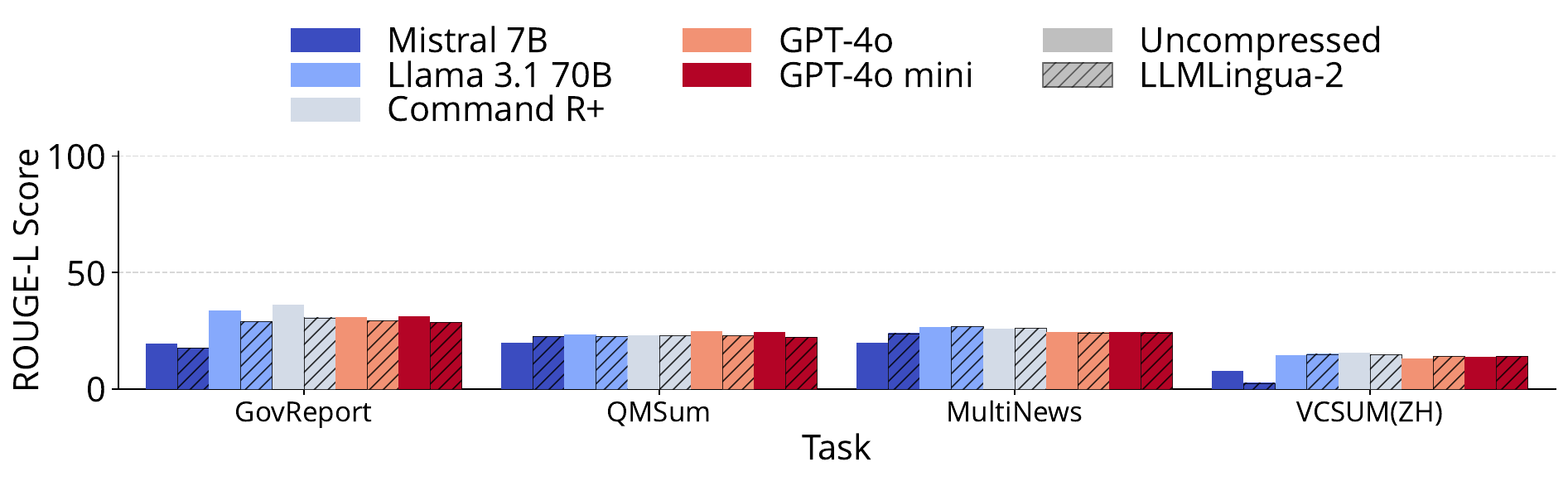}}
    \subfigure[Few Shot Learning]{\includegraphics[width=0.5\linewidth]{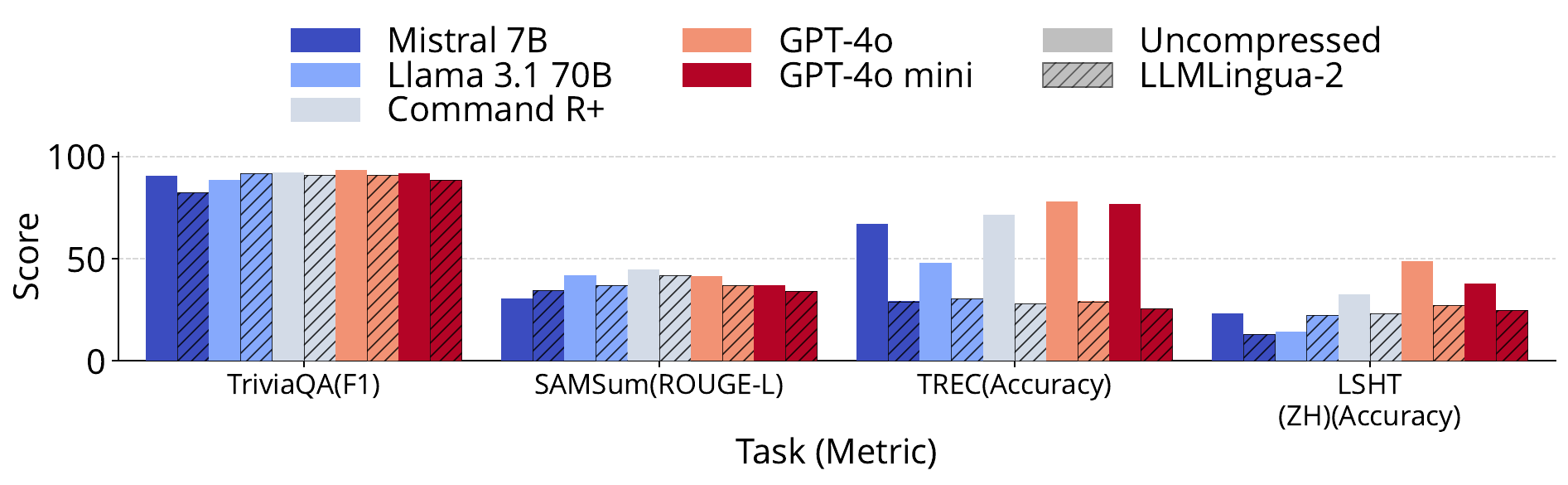}}
    \caption{Response quality of LLMLingua-2 compressed \emph{LongBench} prompts, using different LLMs, compared to the uncompressed baseline dissected into task types.}
    \label{fig:response-quality}
\end{figure*}
\subsection{Response Quality Impact}
We evaluate the impact of prompt compression on response quality using the \emph{LongBench} dataset~\cite{bai_longbench_2024}, extending the original work by Pan et al.~\cite{pan-etal-2024-llmlingua} to a wider set of open-source and proprietary LLMs.
Unlike prior work, we report detailed subtask-level results across all six task categories, revealing significant variability in performance degradation under compression.
All prompts were compressed using LLMLingua-2, which achieved the best latency-performance in our earlier experiments.
We fixed the target prompt length to 3,000 tokens, corresponding to an average compression ratio of $\sim$3.7$\times$ (from 10,300 tokens).
Metrics vary across tasks (indicated in brackets where applicable), though higher scores consistently indicate better performance.
The models were executed using a temperature of $0$, ensuring greedy decoding and therefore, consistent outputs.

\subsubsection{Code Generation.} 
As shown in Figure~\ref{fig:response-quality}a, the code-related tasks exhibit minimal performance degradation on the surface.
However, closer inspection reveals low compression ratios (e.g., $1.4\times$ for the \emph{LongBench} LCC task), which limit both compression benefit and potential speed-up.
Moreover, significant drops in edit similarity suggest that even slight compression harms model outputs.
For RepoBench-P, performance appears stable under compression, but this is misleading: only code snippets are compressed, not the code to be completed.
Zero-shot evaluation confirms that code completion remains robust even with missing examples.
Overall, prompt compression adds latency without quality benefits for code generation.

\subsubsection{Single- and Multi-Document QA.}
Figure~\ref{fig:response-quality}b shows varying model responses to compression.
Mistral 7B maintains or slightly improves performance, likely because its 8,000-token context limit leads to truncation for tasks like NarrativeQA (avg.\ 30,000 tokens).
Compression allows the full context to fit, improving relevance.
In contrast, larger models with longer context windows show performance drops under compression.
For the Chinese MultiFieldQA (ZH) task, performance degrades to roughly 75\% of the original, showing that LLMLingua-2 generalizes moderately to other languages.
Compression results for multi-document QA tasks are similar to single-document QA but with smaller losses—or even slight improvements—especially for Mistral 7B (Figure~\ref{fig:response-quality}c).
Again, context truncation in the original prompt (e.g., HotPotQA, 2WikiMultihopQA, MuSiQue) limits model access to relevant information.
Compression enables fuller context processing, which can enhance accuracy.

\subsubsection{Synthetic Tasks.} 
Synthetic tasks exhibit the strongest negative effects (Figure~\ref{fig:response-quality}d).
Passage Counting accuracy drops from below 20\% to below 4.5\% after compression, likely due to disrupted passage structure.
For Passage Retrieval, models like LLaMA 3.1 70B and GPT-4o initially perform well, but compression reduces accuracy—sometimes to below 50\% because it depends on the structural cue of paragraph numbering, which is lost through compression.
The findings indicate that LLMLingua-2 is unsuitable for compressing structured synthetic datasets.

\subsubsection{Summarization.}
For summarization tasks, performance remains stable under high compression ratios—up to $5.7\times$ (Figure~\ref{fig:response-quality}e).
This suggests that summarization is robust to input reduction. An exception is MultiNews, where the average input length is only 2,600 tokens, leading to an average compression ratio of just $1.26\times$, with many samples uncompressed.
Consequently, this task is not suitable for evaluating compression efficacy under our setup.

\subsubsection{Few-Shot Learning.} 
Few-shot tasks involve conditioning via a small number of examples, which were compressed to fit within the target token limit, while the actual task inputs remained unchanged.
For TriviaQA and SAMSum, compression causes only minor degradation, and zero-shot performance matches the baseline—indicating the examples were not necessary.
Thus, zero-shot prompting is preferable to avoid compression overhead. In contrast, TREC and LSHT-ZH—both classification tasks—suffer performance drops up to 52\%.
The compression process likely removes class-indicative patterns from the examples, making them unsuitable for model inference.

\subsubsection{Summary.} 
GPT-4o and LLaMA 3.1 70B perform best across tasks.
However, the ability of LLMLingua-2 to preserve response quality under compression varies significantly.
Tasks such as summarization remain unaffected even under high compression, while others, such as passage retrieval or few-shot classification, fail due to structural dependencies or pruned class indicators.
A third category includes tasks where compression is unnecessary, as zero-shot baselines already perform well.
Compression proves most useful when prompt length exceeds the model's context window, enabling full input access without truncation.

\begin{figure}[tb]
\centering
    \includegraphics[width=0.55\linewidth]{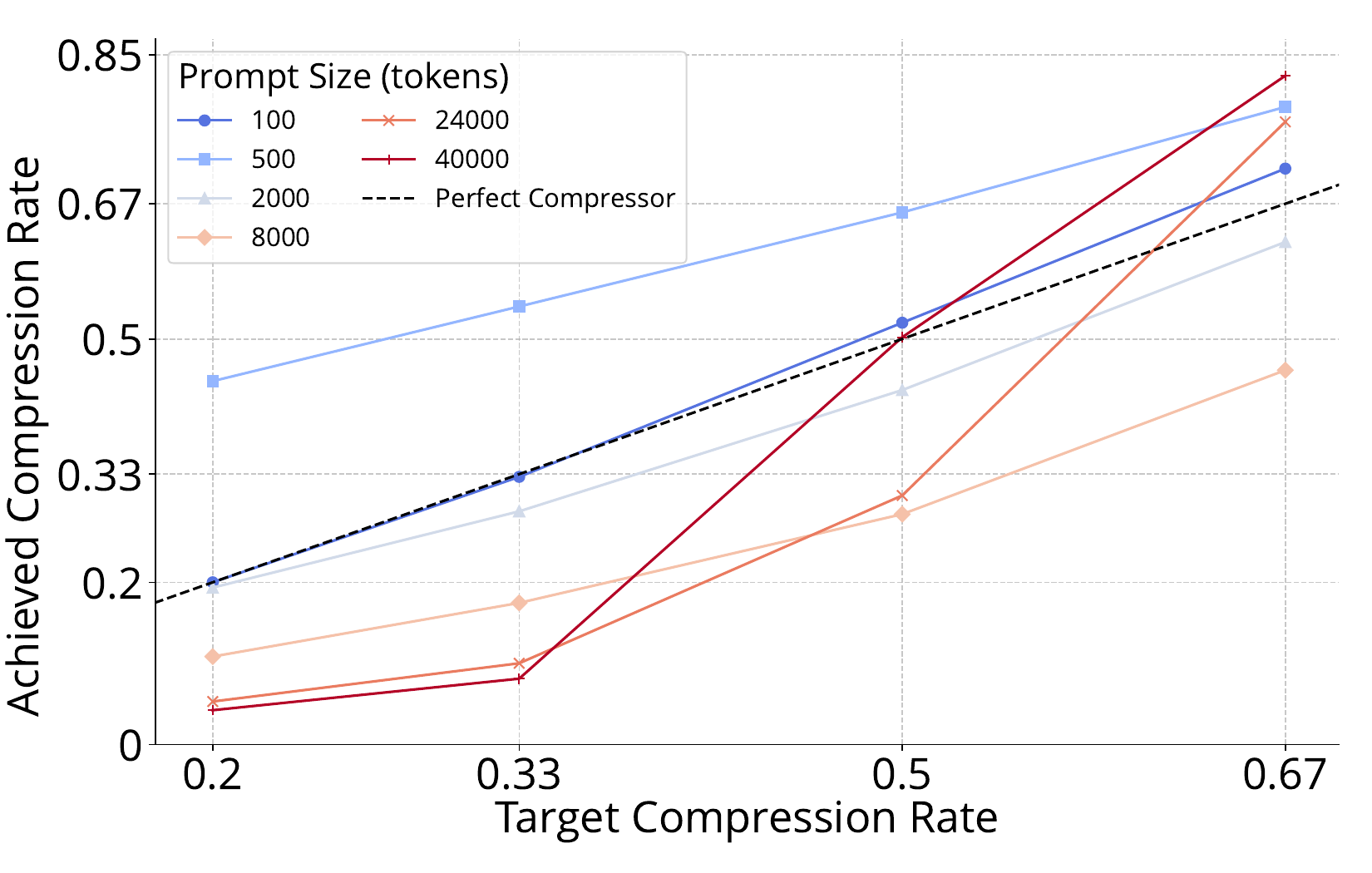}
    \caption{Target compression rate adherence for LLMLingua in dependence on prompt length, compared to a perfect compression, which matches the given compression rate. The compression model does not achieve the given compression rate, which leads to  unpredictable API costs, latency and quality.}
    \label{fig:rate-adherence}
\end{figure}
\subsection{Compression Rate Adherence}
For predictable latency and cost reductions, adherence to the target compression rate is essential.
Therefore, we evaluated the achieved compression rates for LLMLingua and LLMLingua-2 across different prompt sizes and observed substantial differences between the two approaches (see Figure~\ref{fig:rate-adherence}).
LLMLingua generally fails to maintain consistent compression rates, whereas LLMLingua-2 reliably adheres to the specified target.
For LLMLingua, the best adherence occurs at a prompt length of 100 tokens, with a mean absolute error below 0.05.
However, notable deviations appear at prompt lengths of 500, 8,000, and 40,000 tokens.
At 500 tokens, the discrepancy occurs because this length is not a multiple of the in LLMLingua included ITPC algorithm’s iteration size (see \cite{jiang-etal-2023-llmlingua}), leaving the final segment uncompressed and distorting the compression result.
For prompts exceeding 8,000 tokens, the error surpasses 0.15, indicating difficulty in accurately estimating token counts for longer sequences.
These inconsistencies reduce predictability in latency and cost savings.
In contrast, LLMLingua-2 resolves this issue and maintains tight adherence to the target rate across all prompt lengths tested.
\subsection{GPU Memory Requirements}
Prompt compression also raises questions about hardware suitability, particularly regarding memory requirements for predictable deployment at scale.
We measured the memory consumption of both LLMLingua and LLMLingua-2 (base and small variants) on an Nvidia A100 (40\,GB) across various input lengths.

LLMLingua, based on the LLaMA 2 7B model, requires 12.5\,GB of GPU memory, increasing to 16.5\,GB for prompts up to 8,000 tokens.
This makes LLMLingua incompatible with consumer-grade hardware such as the Nvidia GTX 1080 Ti or Apple M1 Pro processors when processing longer inputs (e.g., $\geq$4K tokens).
In contrast, the LLMLingua-small variant, built on GPT-2, requires only a fraction of this memory—reaching a maximum of 1.2\,GB for 1,000-token prompts, which corresponds to the model’s full context capacity.
LLMLingua-2 introduces a more efficient memory footprint.
The base variant starts at approximately 2\,GB and increases to 3.25\,GB for 48K-token prompts.
The small variant is even more lightweight, requiring 0.66\,GB for short inputs and up to 1.5\,GB for maximum-length prompts.
All LLMLingua-2 variants were evaluated using a default batch size of 50.
Although this setting does not fully utilize high-end hardware like the A100, it ensures comparability across devices.
In practice, the batch size can be increased to optimize GPU utilization further.

\section{Conclusion}
\label{sec:conclusion}
We evaluated the LLMLingua prompt compression family with a focus on latency reduction using self-hosted target models across different hardware platforms.
In realistic scenarios involving optimized inference frameworks, we found no substantial improvement in end-to-end latency using non-batched target model execution. 
Speed-ups greater than $1.3\times$ were only observed when using non-optimized frameworks, high-end GPUs such as the Nvidia A100, or very long prompts exceeding 10,000 tokens combined with compression rates above $4\times$.
Generally, the maximum speed-up achievable using a token-pruning prompt compression technique is limited to acceleration of the prefill phase and the reduction of the compression latency overhead. 
Among all evaluated variants, only LLMLingua-2 proved practical under these conditions. Other LLMLingua versions exhibited rate-dependent latency, high memory requirements, and poor adherence to the target compression ratio. 
Response quality varied significantly across tasks: summarization and question answering remained robust under compression, whereas code completion, synthetic tasks, and few-shot examples were sensitive or unsuitable -- especially when zero-shot performance was already comparable.

\begin{credits}
\subsubsection{\ackname} This work is supported by the German Federal Ministry of
Education and Research (BMBF, SCADS22B) and the Saxon
(2023) State Ministry for Science, Culture and Tourism (SMWK)
by funding the competence center for Big Data and AI "ScaDS.AI Dresden/Leipzig“.

\subsubsection{\discintname}
The authors have no competing interests to declare that are relevant to the content of this article. 
\end{credits}

\bibliographystyle{splncs04}
\bibliography{mybibfile2}

\end{document}